\documentclass[a4paper]{article}
\usepackage[utf8]{inputenc}
\usepackage[T1]{fontenc}
\usepackage{graphicx}
\usepackage{grffile}
\usepackage{longtable}
\usepackage{wrapfig}
\usepackage{rotating}
\usepackage[normalem]{ulem}
\usepackage{amsmath}
\usepackage{textcomp}
\usepackage{amssymb}
\usepackage{capt-of}
\usepackage{hyperref}
\usepackage{subfigure}
\usepackage[spanish, english]{babel} \usepackage[utf8]{inputenc}
\usepackage{authblk}
\author[1]{L.A.Rodriguez}
\affil[1]{Dpto. Física, Fac. Cs. Exactas, Nat. y A., Univ. Nacional del Nordeste,Ctes, Argentina}
\author[2]{ H.C. Achitte Schmutzler}
\affil[2]{Dpto. Biología, Fac. Cs. Exactas, Nat. y A., Univ. Nacional del Nordeste,Ctes, Argentina}
\author[2]{ M.I. Dufek}
\author[1]{ Guillermo P.Ortiz}
\author[3]{ W. Luis Mochán}
\affil[3]{Instituto Cs. Físicas, Univ. Nacional Autónoma de México, Cuer., Mor., México}
\date{\today}
\title{Anisotropic optical response of arthopods's cuticle}
\hypersetup{
 pdfauthor={Guillermo Ortiz},
 pdftitle={Anisotropic optical response of arthopods's cuticle},
 pdfkeywords={},
 pdfsubject={},
 pdfcreator={Emacs 27.1 (Org mode 9.3)},
 pdflang={English}}
\begin{document}

\maketitle
\begin{abstract}

The structure of arthropods cuticle consists of layers of
microfilamentary chitin particles. The layers are stacked one on top
of the other performing an helical (Bouligand helix) pattern. This
cuticle structure generate optical phenomena such as structural color
and birefringence that develop as metalic appearance or iridescence
and polarized reflectance or optical activity. We model the
anisotropic optical responses of arthropods cuticle using \texttt{Photonic}
package to obtain the macroscopic dielectric tensor of a constitutive
layer in the Bouligand helix. We found the anisotropy of that
dielectric tensor depends on chitin particle shape, reticular
arrangement and filling fractions. As a main result, we obtained a
cuticle model with tunable reflectance band gaps that are very
sensitive to such geometrical parameters and to \(\theta\) angle that
controls the helix pitch.  We can explain structural color based on
the reflectance band gap introduced by anisotropy and its shifts
mediated by \(\theta\) instead of as usual approach that consider
multiple arrangement of layers pairs with different thickness and
optical properties.
\end{abstract}
\section{Introduction}
\label{sec:orgc593fd4}

There are several structured cuticle that behave like complex photonic
crystals generating optical phenomena in sclerotized integument of
arthropods like structural color, luminescence, ultraviolet signals,
polarized reflectance, and depolarization
\cite{Neville(1969),Parker(1998),Galusha(2008),Sharma(2009)}

The basic model of a stratified cuticle arranged parallel in
multilayered structure that can alternates high and low refractive
indices, i.e. a Bragg mirror, was broadly used to
explain\cite{Neville(1975),Luna(2013),Gullan(2014)} colorful
iridescent bright reflection due to constructive/destructive
wavelength superposition by interference phenomena. Also, sections of
samples of the dorsal cuticles analyzed with SEM and TEM images
confirms the evidence of cuticle structure of the stratified type in
several different arthropods \cite{Parker(2000),Lenau(2008)}. Of
course, a Bragg mirrors arrangement only does not can explain more
specific optical responses such as the optical activity. When the
circular polarized light rotates following the same counterclockwise,
it is called right circular polarization and otherwise is left
circular polarization \cite{Neville(1993)}. Several beetle species reflect
circular polarized light also the marine stomatopod odontodactylus
scyllarus and several species of firefly \cite{Pye(2010),Carter(2016)}

The typical structure of the cuticle in arthropods consists of layers
of chitin microfilamentary particles into protein matrix
\cite{Lenau(2008)}.  The chitin microfilaments are arranged parallel
and into helical layers throughout the cuticle behaving similarly to
liquids crystals in cholesteric
phases\cite{Neville(1969),Bouligand(1972)}, it means that particles
have a same orientation in a thin layer but stratified and
progressively changing its orientation rotating around the layer
normal directions deploying an helix structure. The helical model
proposed by Bouligand\cite{Bouligand(1972)}, was assumed
\cite{Neville(1969)} and later demonstrated
\cite{Jewell(2007),Sharma(2009)} that the cuticle is composed of layer
with a successive, counterclockwise helical rotation of suitable pitch,
i.e. the length traveled along the helix axes for a full rotation
around such axes. Due to elongated shape of chitin
particles in a layer it has a \emph{form} birefringence properties. The
birefringence is the optical property of a material that has different
refractive indices to light rays. Crystals of calcite are a very well
known example displaying this property. Due to axial symmetry in the
molecular arrangement into the crystalline structure the birefringence
of the calcite is an intrinsic property\cite{Hecht(1986)}.  The form
birefringence is an important feature of the individual layers, as it
contributes to the \emph{strength} of the helix\cite{Carter(2016)}. It is
related to the anisotropy that could be due to the aspect ratio of
chitin microfilaments or more accurately due to depolarization factor
that depend on particle geometry shape and materials. For large
differences from one to another of the depolarization factor
corresponding to two of its principal axis, larger is the
anisotropy, and thus layers constitutes to an helix structures
optically stronger.

The Bouligand structures served as the basis for explaining the chitin
structure and organization, also to estimate the birefringence and
anisotropy from many crustaceans and insects. The polarization of the
cuticle structure in insect was studied and looked as a simple model
of the reflection mechanism \cite{Carter(2016)}. To known the possible
structural origins of the reflection spectra of circular polarized
light in beetle cuticle in Ref.\cite{Carter(2016)} is compared the
observed spectra in beetle cuticle with spectra modeled as a
multilayer transfer matrix method using the Birefringent Thin Films
Toolbox (BTFT) \cite{McCall(2015)}. In the BTFT method the form
birefringence is estimated through a proposed structure function that
perform a mix of voids and chitin’s particles with cylindrical shape
using the Bragg-Pippard analytical formula\cite{Oldenbourg(1989)} to
approximate the macroscopic dielectric permittivity of chitin
microfilaments composite media. In that model the authors introduced
some parameters determined by fitting against data obtained from
reflectance measurements which means that they do not model the
anisotropy indeed.

In this work, we propose to obtain the anisotropic optical response of
a chitin microfilaments layer by calculating their macroscopic
dielectric tensor from which and using matrix transfer method we get
the reflectance of the cuticle structure. Thanks to this procedure we
can explain the reflectance features of the structural color observed
in beetles like in Ref.\cite{Luna(2013),Parker(2000)} in another
way. The reflectance band gap and also its shifts are produced by
anisotropy that is mediated by microstructure details and the helix
pitch instead of the often approach considering multiple arrangement
of layers pairs with different optical properties and thickness.

We organize this work introducing in Sec.\ref{orgf2a5519} the theory and
methods to describe calculation of the macroscopic dielectric tensor
for a layer of chitin microfilaments particle. In Subsec.\ref{orgb5770a2} we
resume an appropriate transfer matrix method that we used in
Subsec:\ref{orge860b87} to obtain the cuticle anisotopic reflectance. In
Sec.\ref{org98d0d52} we present and discuss the results we found, and
Sec.\ref{org98cc6b2} is devoted to our conclusion.

\section{Theory and Methods \label{orgf2a5519}}
\label{sec:orgb8f3b63}

We want to model the optical response of a layered system with total
thickness \(L\) where each layer is constituted by intertwined chitin
macromolecules developing a planar array of elongated microfilamentary
particles with cross section of a few nm in diameters. In these layers
exists a preferential direction defined by the alignment in average of
the major axes of the microfilamentary particles. However, each layer has a
different average alignment direction and thus we chose the laboratory
system reference to describe the optical response of the whole layered
structure. We model the optical response of each layer using
\texttt{Photonic} and employing a transfer matrix method to obtain the
reflectance R and transmittance T of the total system.

\begin{figure}[htbp]
\centering
\includegraphics[width=0.6\textwidth]{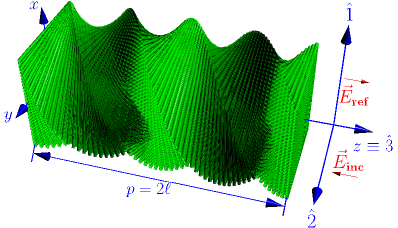}
\caption{\label{fig:org6b9b489}Bouligand structure scheme for \(M=73\) stacked layers progressive rotated each in \(\theta=5\) around \(z\) axes in the laboratory reference frame \(x\),\(y\),\(z\) with helix pitch \(p\) and period \(\ell\). Fields \(\vec E_{\mbox{inc}}\) and \(\vec E_{\mbox{ref}}\) for an incident probe beam and its reflection in the \(z\) axes direction of the laboratory reference and equal to axes \(\hat 3\) belonging to body frame reference with principal axis \(\hat1\) and \(\hat2\).}
\end{figure}

Fig. \ref{fig:org6b9b489} displays an scheme of the progressive rotated \(M\) layers
developing an helical Bouligand structure \cite{Bouligand(1972)}.
Layers are stacked and successively rotated  counterclockwise in \(\theta\).
They are defined for a simple example with thirty particles of
cylindrical shape with aspect ratio height to diameter of thirty
arranged side by side forming a squared base. The helix pitch is \(p\)
and due to a half of full rotation symmetry of cylindrical particles around the
normal to its axes the Bouligand structure has a period \(\ell=p/2\).
Normal incidence of a probe beam is taken at normal direction of
layers which named \(\hat z\) in the laboratory frame (\(\hat x\),\(\hat
y\),\(\hat z\)) and equivalent to axes \(\hat 3\) in the body reference
frame (\(\hat1\),\(\hat2\),\(\hat3\)).

\begin{figure}[th]
     \centering
     \begin{subfigure}
    \centering
        \includegraphics[width=0.35\textwidth]{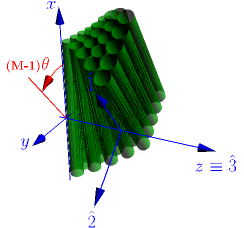}
     \end{subfigure}
     \begin{subfigure}
    \centering
        \includegraphics[width=0.35\textwidth]{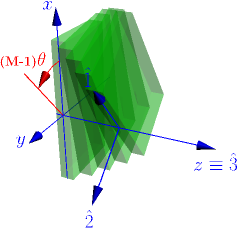}
     \end{subfigure}
\caption[\texttt{Photonic}]{\label{org5f27c25}Left: Details of Fig.\ref{fig:org6b9b489} for cylindrical chitin particles. Right: Scheme substituting particles specific geometry with homogeneous slabs that allows us to model anisotropic response calculated with \texttt{Photonic}. There are principals axis in layers planes \(\hat1\),  \(\hat2\) and the respective laboratory axis \(x\), \(y\). The \(z\) and \(\hat3\) axis are the same. The progressive rotation \((M-1)\theta\) follows for each of \(1\ldots M\) layers.}
\end{figure}

Fig. \ref{org5f27c25} displays at the left a zoom of Fig. \ref{fig:org6b9b489} indicating some \(M\)
layers of cylinders progressively rotatated in \(\theta\). In the right
part of Fig. \ref{org5f27c25} the \(M\) layers of cylinders are substituted by
\(M\) homogeneous slabs corresponding each to the macroscopic dielectric
tensor obtained using \texttt{Photonic} \cite{Mochan(2016)}. As expected, it
is diagonal in the body frame (\(\hat1\),\(\hat2\),\(\hat3\)) and we write the
parallel part
\begin{equation}\label{eq:epar_prin}
\epsilon^M_{\parallel}=\left(\begin{array}{cc}\epsilon_{11}&0\\0&\epsilon_{22}\end{array}\right)
\end{equation}
As usual, we use a linear transformation of rotation in the layer
plane to obtain \(\epsilon_{\parallel}\) in the laboratory frame
\begin{equation}\label{eq:epar}
\epsilon^M_{\parallel}=\left(\begin{array}{cc}\epsilon_{xx}&\epsilon_{xy}\\\epsilon_{yx}&\epsilon_{yy}\end{array}\right),
\end{equation}
and it is always symmetric \(\epsilon_{xy}=\epsilon_{yx}\).
\begin{figure}[htbp]
\centering
\includegraphics[trim='2cm 2.5cm 2cm 2.5cm',width=0.4\textwidth]{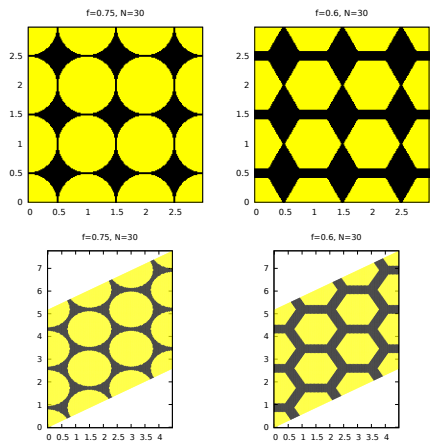}
\caption[ ]{\label{fig:org57096f7}Upper (Lower) panels shows tiled 3 \texttimes{} 3 times square (triangular) unit cells cuts for  \((\hat2,\hat3)\) frame body directions. In the left (right) for cylindrical (hexagonal) particles sections for \(f=0.75\) (\(f=0.6\)) filling fractions. There are \(2N+1\) points by unit cells side used for the particle geometry definition.}
\end{figure} Fig. \ref{fig:org57096f7} displays the geometries and the
particles section views with the lattice unit cells we have explored.
Note that for clearness in the schema of the cuticle structure we put
only one cylinder stack in each layer shows in Fig.\ref{org5f27c25} left
panel. But, we expect that there could be a bunch of chitin particles
more or less aligned and that units cells cuts displayed in Fig.\ref{fig:org57096f7}
allows a good enough representation of each layer for the calculation
of macroscopic dielectric tensor responses.  Fig.\ref{fig:org57096f7} displays in
upper (lower) panel square (triangular) unit cells \((\hat2,\hat3)\)
frame body cuts directions tiled 3 \texttimes{} 3 times.  In the left
(right) panel of Fig.\ref{fig:org57096f7} for cylindrical (hexagonal) particles
section with \(f=0.75\) (\(f=0.6\)) approximately to the maximum filling
fractions for the square lattice. Note that for triangular lattices a
larger \(f\) values are allowed due to the smaller unit cell surface in
such cases.  With numerical computation via \texttt{Photonic} there are
parameter \(N\) and \(N_h\) that need to be tested for witness of
convergence. The \(2N+1\) point per unit cells side are used to define
the particle geometry, while \(N_h\) define the number of states used to
represent physical objects in the Fourier space for Haydock recursion
method\cite{Mochan(2016)}.

From the self-consistent fields obeying Maxwell Eqs. we may obtain the
optical response of the layered system. We first suppose that exist
two waves inside of each layer with wavevectors \(\vec
k=\pm\frac{\omega}{c}n \hat z\), with \(n\) depending on dielectric
tensor components. Note that for isotropic materials
\(n=\sqrt{\epsilon}\) with \(\mu=1\) and \(\epsilon\) the permittivity
media.  For anisotropic media \(n\) is obtained from the dispersion
relation.  Alternatively, we may proceed applying Faraday and
Ampere-Maxwell Eqs. in the laboratory frame reference to obtain from
\(\vec E\) and \(\vec H\) the named \emph{continuous vector}
\((E_x,H_y,E_y,H_x)\) \cite{McCall(2015)} corresponding to parallels
components \(\vec E_{\parallel}\) and \(\vec H_{\parallel}\) to the layers
interfaces, which must obeys
\begin{equation}\label{eq:eigen}
\left(\begin{array}{cccc}0 &1 &0 &0\\\epsilon_{xx}&0 &\epsilon_{xy}&0\\0&0&0&-1\\
-\epsilon_{xy}&0&-\epsilon_{yy}&0\end{array}\right)\left(\begin{array}{c}
E_x\\H_y\\E_y\\H_x\end{array}\right)=n\left(\begin{array}{c}
E_x\\H_y\\E_y\\H_x\end{array}\right)
\end{equation}
eigenvalues and eigenvectors equations. The characteristic polynomial
give us 4 solutions for \(n\)
\begin{equation}\label{eq:n}
n^2=\frac{\epsilon_{xx}+\epsilon_{yy}}{2}\pm\frac{1}{2}\sqrt{(\epsilon_{xx}-\epsilon_{yy})^2+4\epsilon_{xy}^2}
\end{equation}
which we may easily identify as is diagonal case
with solutions \(n_1^{\pm}=\pm\sqrt{\epsilon_{11}}\) and
\(n_2^{\pm}=\pm\sqrt{\epsilon_{22}}\) corresponding to the two sets of counter
propagating waves normal modes of each layer.

\subsection{Tranfer matrix \(\mathbf{M}\) for anisotropic system \label{orgb5770a2}}
\label{sec:org0f287bc}

Here we briefly resume well known theory for propagating field Maxwell
Eqs. solutions in stratified media \cite{BornWolf(1999)}, but instead
to follow the backward fields transfer description
\cite{McCall(2015),BornWolf(1999)} we will describe a forward fields
transfer like in Ref. \cite{Missoni(2020),Puente(2020)}, because the
last ones follows straightforwardly. From \(\vec E_{\parallel}\) and
\(\vec H_{\parallel}\) we write the continuos vector in terms of the
eigenvectors \(\left\{\vec v_1^+,\vec v_1^-,\vec v_2^+,\vec
v_2^-\right\}\), as
\begin{equation}\label{eq:cl}
\left(\begin{array}{c}
E_x\\H_y\\E_y\\H_x
\end{array}\right)=
\gamma_1^+\vec v_1^+ +\gamma_1^-\vec v_1^- +\gamma_2^+\vec v_2^+ +\gamma_2^-\vec v_2^-
\end{equation}
using the complex amplitudes \(\gamma_1^{\pm}\) and
\(\gamma_2^{\pm}\). Introducing matrix \(\mathbf B\) made up with columns
of that eigenvectors, thus \(\mathbf B\) only depends on the layer
permittivity tensor components. In any plane \(z\) inside each layer,
the amplitudes from laboratory frame can be obtained
\begin{equation}\label{eq:gamma_z}
\left(\begin{array}{c}
\gamma_1^+\\\gamma_1^-\\\gamma_2^+\\\gamma_2^-
\end{array}\right)_z=
\mathbf B^{-1}
\left(\begin{array}{c}
E_x\\H_y\\E_y\\H_x
\end{array}\right)_z.
\end{equation}
Note from \eqref{eq:cl} that we can assign to the amplitudes of each
eigenvectors the phase \(\pm k_0 n_iz\) with \(i=1,2\) and \(k_0=\omega/c\)
as function of the \(z\) position propagation coordinates, thus we may
write for \(z_1<z<z_2\)
\begin{equation}\label{eq:gamma_prop}
\left(\begin{array}{c}
\gamma_1^+\\\gamma_1^-\\\gamma_2^+\\\gamma_2^-
\end{array}\right)_{z=z_2}=
\mathbf{P}(z_2-z_1)
\left(\begin{array}{c}
\gamma_1^+\\\gamma_1^-\\\gamma_2^+\\\gamma_2^-
\end{array}\right)_{z=z_1}.
\end{equation}
where
\begin{equation}\label{eq:P}
\mathbf{P}(z)=
\left(\begin{array}{cccc}
e^{ik_0n_1z} &0&0&0\\
0&e^{-ik_0n_1z}&0&0\\
0&0&e^{ik_0n_2z}&0\\
0&0&0&e^{-ik_0n_2z}
\end{array}\right)
\end{equation}
Using \eqref{eq:gamma_prop} and \eqref{eq:gamma_z} we obtain the
continuos vector transformation from \(z=z_1\) to \(z=z_2\) positions
\begin{equation}\label{eq:EH_z2}
\left(\begin{array}{c}
E_x\\H_y\\E_y\\H_x
\end{array}\right)_{z=z_2}
=
\mathbf B
\mathbf{P}(z_2-z_1)
\mathbf B^{-1}
\left(\begin{array}{c}
E_x\\H_y\\E_y\\H_x
\end{array}\right)_{z=z_1}
\end{equation}
For a layer of thickness \(d\) with its left interface at \(z=z_0\),
 \(\mathbf{M}(d)=\mathbf{B} \mathbf{P}(d)\mathbf{B}^{-1}\) allows us to
 obtain the transformation of continuos vector from \(z=z_0\) to
 \(z=z_0+d\)
\begin{equation}\label{eq:M}
\left(\begin{array}{c}
E_x\\H_y\\E_y\\H_x
\end{array}\right)_{z=z_0+d}
=
\mathbf M(d)
\left(\begin{array}{c}
E_x\\H_y\\E_y\\H_x
\end{array}\right)_{z=z_0}.
\end{equation}
Then, for a layered system with thickness \(L\) like in Fig.\ref{org5f27c25}
with its first interface in \(z=0\), and with \(N\) contiguous layers each
one of width \(d_i\) and \(i=1\ldots N\), we obtain
\begin{equation}\label{eq:EH_L}
\left(\begin{array}{c}
E_x\\H_y\\E_y\\H_x
\end{array}\right)_{z=L}
=
\mathbf M
\left(\begin{array}{c}
E_x\\H_y\\E_y\\H_x
\end{array}\right)_{z=0}
\end{equation}
where \(L=\sum_i^Nd_i\) and \(\mathbf{M}=\mathbf{M_N}(d_N)
\mathbf{M_{N-1}}(d_{N-1})\ldots \mathbf{M_1}(d_1)\) defines the \emph{transfer
matrix} of the whole system.

\subsection{Reflection coefficients for anisotropic system \label{orge860b87}}
\label{sec:orgf5de2f5}

Note that in \eqref{eq:EH_L} we have the total parallel field
components to the interfaces at \(z=0\) and at \(z=L\). Thus, for the
incident and reflected fields at \(z=0\) and for the transmitted field
at \(z=L\), we can write
\begin{equation}\label{eq:EH_T_L}
\left(\begin{array}{c}
E_{tx}\\H_{ty}\\E_{ty}\\H_{tx}
\end{array}\right)_{z=L}
=
\mathbf M
\left(\begin{array}{c}
E_{0x}+E_{rx}\\H_{0y}+H_{ry}\\E_{0y}+E_{ry}\\H_{0x}+H_{rx}
\end{array}\right)_{z=0}.
\end{equation}
Also, note that as the incident and trasmitted media are isotropic, we
have \(H_{tx}=E_{ty}/Z_s\) and \(H_{ty}=E_{tx}/Z_s\) at \(z=0\), while
\(H_{0x}+H_{rx}=(E_{0y}-E_{ry})/Z_0\) and
\(H_{0y}+H_{ry}=(E_{0x}-E_{rx})/Z_0\) at \(z=L\), where \(Z_i=1/n_i\) are
impedance for normal incidence and \(n_0\), \(n_s\) refractive index for
incident medium and substrate medium. Such relations allows us to
equate a pair of transmitted fields expressions to reduce from 4\texttimes{}4
to one of 2\texttimes{}2 Eqs. system. After a simple algebra and rearranging
terms, we can write
\begin{equation}\label{eq:E_2x2}
\mathbf M_1
\left(\begin{array}{c}
E_{0x}+E_{rx}\\(E_{0x}-E_{rx})/Z_0
\end{array}\right)
=
\mathbf M_2
\left(\begin{array}{c}
E_{0y}+E_{ry}\\(E_{0y}-E_{ry})/Z_0
\end{array}\right),
\end{equation}
where
\begin{equation}\label{eq:M1}
\mathbf M_1
=
\left(\begin{array}{cc}
m_{11}-Z_S m_{21} & m_{12}-Z_S m_{22} \\
m_{31}-Z_S m_{41} & m_{32}-Z_S m_{42}
\end{array}\right),
\end{equation}
and
\begin{equation}\label{eq:M2}
\mathbf M_2
=
\left(\begin{array}{cc}
Z_S m_{23}-m_{13} & Z_S m_{24}-m_{14}\\
Z_S m_{43}-m_{33} & Z_S m_{44}-m_{34}
\end{array}\right),
\end{equation}
with \(m_{ij}\) defined in \eqref{eq:EH_L}. Finally, reflected fields are
\begin{equation}\label{eq:Er}
\left(\begin{array}{c}
E_{rx}\\E_{ry}
\end{array}\right)
=
\mathbf A^{-1}
\left[\begin{array}{cc}
E_{0x}\mathbf M_1-E_{0y}\mathbf M_2
\end{array}\right]
\left(\begin{array}{c}
1     \\1/Z_0
\end{array}\right),
\end{equation}
with 2\texttimes{}2 matrix
\begin{equation}\label{eq:A}
\mathbf A
=
\left[
-\mathbf M_1
\left(\begin{array}{c}
1     \\-1/Z_0
\end{array}\right)
\,\,
\mathbf M_2
\left(\begin{array}{c}
1     \\-1/Z_0
\end{array}\right)
\right]
\end{equation}
From \eqref{eq:Er} we can obtain the useful anisotropic reflectance
coefficients \cite{Yeh(2005)}
\begin{eqnarray}\label{eq:rxy}
r_{xx}&=&E_{rx}/E_{0x}\quad (E_{0y}=0)\\\nonumber
r_{xy}&=&E_{rx}/E_{0y}\quad (E_{0x}=0)\\\nonumber
r_{yy}&=&E_{ry}/E_{0y}\quad (E_{0x}=0)\\\nonumber
r_{yx}&=&E_{ry}/E_{0x}\quad (E_{0y}=0)
\end{eqnarray}

\section{Results \label{org98d0d52}}
\label{sec:orge92ebdc}
\begin{figure}[th]
     \centering
     \begin{subfigure}
    \centering
        \includegraphics[width=0.6\textwidth]{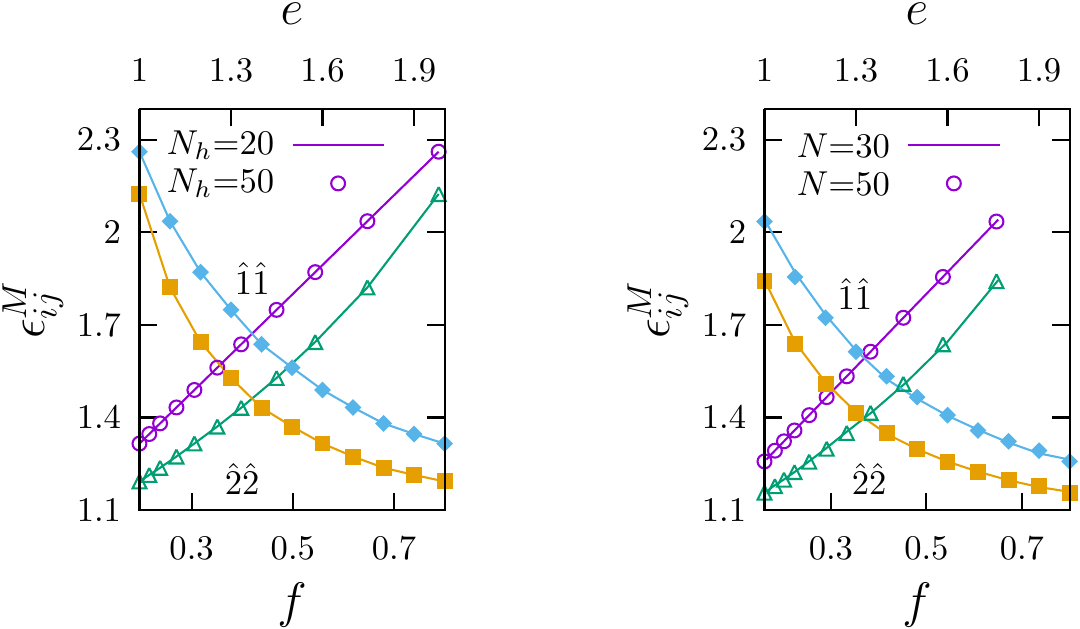}
     \end{subfigure}

     \begin{subfigure}
    \centering
        \includegraphics[width=0.6\textwidth]{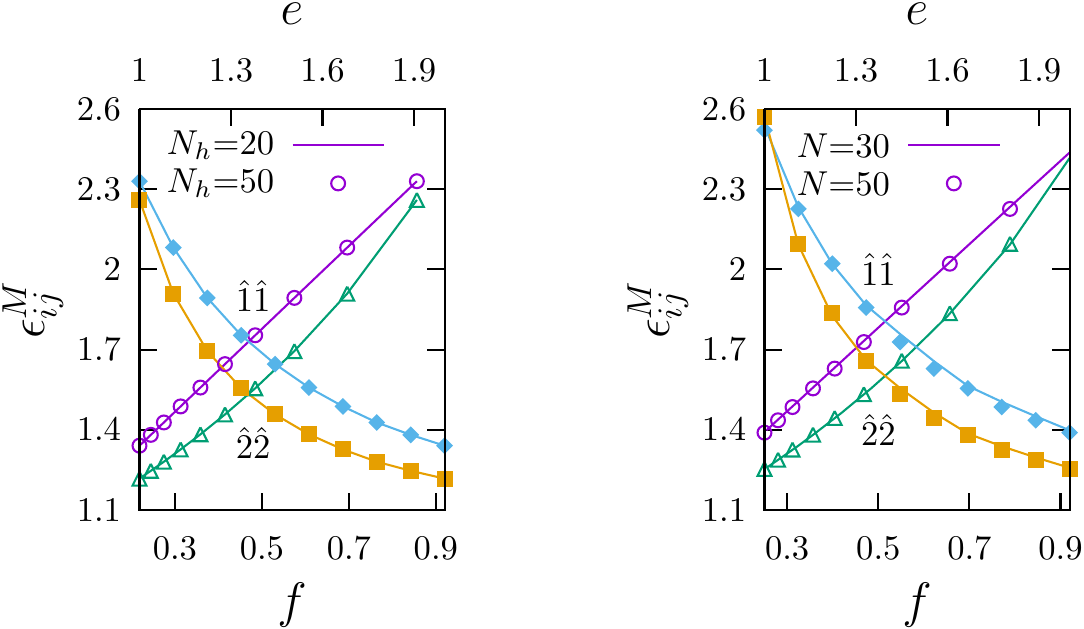}
     \end{subfigure}
\caption{\label{org40b0b6f}The diagonal macroscopic tensor components \(\epsilon^M_{ij}\) \(i=j=1,2\) versus \(f\) and the  aspect ratio \(e\) for left (right) cylinder (hexagonal) Eq.\eqref{eq:epar_prin}. Upper (Lower) panel square (triangular) lattice cell cases. Linespoints versus continuous line are used to test \(N\) and \(N_h\) convergences with no appreciable differences above \(N=30\) and \(N_h=20\).}
\end{figure}

\begin{figure}[htbp]
\centering
\includegraphics[trim='3cm 4cm 3cm 4cm',width=0.5\textwidth]{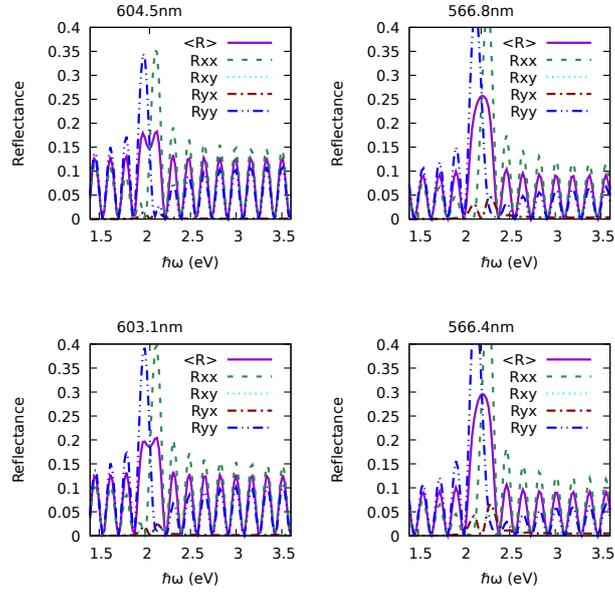}
\caption{\label{fig:org5db2ee0}Reflectance R\textsubscript{ij} versus photon energy \(\hbar\omega\) in eV for Bouligand structure with thickness \(L=\) 2496 nm, \(\theta\) =10° and \(M=\) 216 layers. Cases are ordered as in results of Figs.\ref{fig:org57096f7} and \ref{org40b0b6f}. \(\langle R\rangle_{\text{bgcp}}\) develops at \(\lambda=\) 604.5 (566.8) and 603.1nm (566.4) for cylindrical (hexagonal) particle sections. Continuous line for \(\langle R\rangle= \frac{1}{2} (R_{xx}+R_{yy})\). Dashed lines for \(R_{xx}\), Dashed dotted-dotted  lines for \(R_{yy}\) are one  order of magnitude larger than \(R_{xy}\) and \(R_{yx}\) which are displayed in dotted and dashed-dotted lines.}
\end{figure}

\begin{figure}[htbp]
\centering
\includegraphics[trim='3cm 11cm 3cm 12cm',width=.9\linewidth]{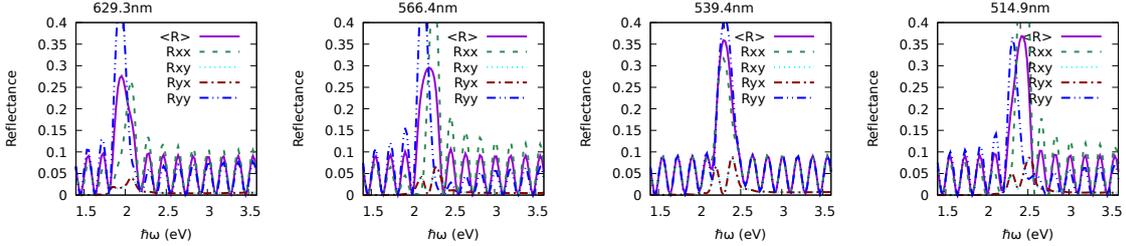}
\caption{\label{fig:org7ed926f}Reflectance R\textsubscript{ij} versus photon energy \(\hbar\omega\) in eV for Bouligand structure similar to Fig.\ref{fig:org5db2ee0} but only for triangular lattice and hexagonal particles in order from left to right with \(\theta=\) 9°,10°,10.5° and 11°. The \(\langle R\rangle_{\text{bgcp}}\) shifts are \(\lambda=\) 629.3, 566.4, 539.4 and 514.9nm, respectively.}
\end{figure}

Fig. \ref{org40b0b6f} displays \(\epsilon^M_{ij}\) with \(i=j=1,2\) versus \(f\) and
the aspect ratio \(e\) corresponding to Eq.\eqref{eq:epar_prin} and
\(e=h/2r\) with particles height \(h\) and radius \(r\). In the upper
(lower) panel for square (triangular) lattice.  Left (Right) panels
corresponds to cylindrical (hexagonal) particle geometry
sections. Convergence of \(\epsilon^M_{ij}\) with \(N\) and \(N_h\) values
used are obtained with no appreciable difference above
\(N=30\) and \(N_h=20\).
Note that differences of \(\epsilon^M_{ii}\) in longitudinal \(\hat 1\hat
1\) respect to transversal \(\hat 2\hat 2\) particles directions
corresponds to the form birefringence we have obtained for each modeled layer
given by the anisotropic diagonal tensor \(\epsilon^M_{ij}\). Fig. \ref{fig:org5db2ee0}
displays anisotropic reflectance
 \(R_{ij}=|r_{ij}|^2\) according to Eq.\eqref{eq:rxy} for the
free standing \((Z_0=Z_s=1)\) Bouligand structure with total thickness
\(L=\) 2496 nm for \(\theta\) =10º and \(M=216\) layers of thickness
\(d=L/M=11.56\) nm each one.  For cylindrical particles section
\(\epsilon^M_{11}=2.2\), \(\epsilon^M_{22}=2.03\) for square lattice (left
upper panel) and \(\epsilon^M_{11}=2.2\), \(\epsilon^M_{22}=2.0\) for
triangular lattice (left lower panel). For hexagonal particles section
\(\epsilon^M_{11}=1.96\), \(\epsilon^M_{22}=1.75\) for square lattice
(right upper panel), and \(\epsilon^M_{11}=1.97\),
\(\epsilon^M_{22}=1.74\) for triangular lattice (right lower
panel). Note that near to \(\hbar\omega_{\text{cyl}}=2.05\) eV for
cylindrical and near to \(\hbar\omega_{\text{hex}}=2.19\) eV for
hexagonal cases we have the centroid of the band gap \(\langle
R\rangle_{\text{bgcp}}\) for the reflectance average \(\langle R
\rangle= \frac{1}{2} (R_{xx}+R_{yy})\) for no polarized light as
expected for the modeled Bouligand structures\cite{Carter(2016)}
\begin{equation}\label{eq:bragg}
1239.8/\hbar\omega=p\frac{\sqrt{\epsilon_{11}}+\sqrt{\epsilon_{22}}}{2}.
\end{equation}
Note that larger values of \(\epsilon^M_{ii}\) for cylindrical case are
also related to the larger \(f\) as well as the larger amplitude of the
oscillation outside the band gap compared to the hexagonal case. Note
also that as a consequence or larger \(\epsilon^M\) the period of such
oscillation is smaller than the respective for hexagonal case. These
facts could explain partially only a larger peaks overlap by
superposition of \(R_{xx}\) and \(R_{yy}\) in the hexagonal compared to
cylindrical case. Also, in Fig. \ref{fig:org5db2ee0} \(\langle
R\rangle_{\text{bgcp}}\) is larger when the anisotropy difference
\(\Delta n=(n_{11}-n_{22})/n_{22}\) increases following \(\Delta n=\)
0.041, 0.049, 0.058 and 0.064 for left upper, left lower, right upper
and right lower panels, respectively. Clearly for \(\Delta n\to\) 0
anisotropy disappear as also any band gap reflectance. The helix
strength related to \(\Delta n > 0\) means that constructive
interference phenomena takes place and for larger strength larger is
the band gap reflectance. Some features of the overlapped peaks depends
on \(\theta\) also. Fig.\ref{fig:org7ed926f} displays
  anisotropic reflectance \(R_{ij}\) for
the similar Bouligand structures of Fig.\ref{fig:org5db2ee0} but now only for
triangular lattice and hexagonal particles with \(\theta=\) 9°,10°,10.5°
and 11°.  Ripples features of the band gap shape related to \(\theta\)
has attracted some attention \cite{Carter(2016)} and would be
explained here. As \(\theta=\) 10° correspond to six full helix rotation
for \(M=216\), a compensated contribution of the orthogonal components
\(R_{xx}\) and \(R_{yy}\) to \(\langle R \rangle\) is obtained as we can
verify from the symmetric reflectance band gap in the second panel of
Fig.\ref{fig:org7ed926f}.  A \(\theta\) value not commensurate with some integer of
full helix rotation will give uncompensated contribution as we can see
in the others panel of Fig.\ref{fig:org7ed926f}.  According to Eq.\eqref{eq:bragg}
and \(p=360 d/\theta\), the \(\hbar\omega\) is proportional to \(\theta\)
as we can verify in Fig.\ref{fig:org7ed926f} for the blue shifted \(\langle
R\rangle_{\text{bgcp}}\) evaluating \(\langle R\rangle\) at
\(\hbar\omega=\) 1.97, 2.19, 2.3 and 2.4 eV corresponding to increasing
\(\theta=\) 9°,10°,10.5° and 11°, respectively but using
\(\lambda=1239.8/\hbar\omega\). Thus, It is interesting to note that
\(\langle R\rangle_{\text{bgcp}}\) changes from \(\lambda_{\text
r}=629.3\) nm at \(\theta=\) 9° to \(\lambda_{\text g}=514.9\) nm at
\(\theta=\) 11° displaying a change from red to green colors with smalls
changes of \(\theta\).

\section{Conclusions \label{org98cc6b2}}
\label{sec:org3d9385f}

In this work we have calculated using \texttt{Photonic}\cite{Mochan(2016)}
the anisotropic macroscopic dielectric tensor of a layer we used to
model the anisotropic reflectance of the Bouligand structures
\cite{Bouligand(1972)} for cuticle made of stratified chitin
microfilamentary particles. We have explored anisotropy of the
macroscopic dielectric tensor taken into account the chitin particles
geometry, lattice arrangement, aspect ratio and the filling
fractions. We have considered the Bouligand cuticle structure that are
characterized by rotated layers of thickness \(d\) developing helical
structure with pitch \(p=360 d/\theta\) and \(\theta\) the rotation angle
between two consecutive layer that constitute the entire cuticle
structure. To take into account the cuticle anisotropic reflectance we
have calculated the dielectric tensor of each rotated layer in the
laboratory frame and we have introduced a transfer matrix method of 4
\texttimes{} 4 components that allows translate self consistently from above
to below the incident electromagnetic fields on the Bouligand
structure.  We found that for no polarized light a reflectance band
gap develops with a centroid position \(\langle R\rangle_{\text{bgcp}}\)
in agreement with the expected constructive interference
Eq.\eqref{eq:bragg} relation. These reflectance band gap can be seen
as the superposition of the two orthogonal responses in the laboratory
frame \(R_{xx}\) and \(R_{yy}\). We show that \(\langle
R\rangle_{\text{bgcp}}\) is larger when \(\Delta n\) increases and that
\(\Delta n\) is very sensitive to the particle geometry and lattice
arrangement. These facts were also noticed by many authors
\cite{Carter(2016)} but anybody overcame the challenge to calculate
cuticle optical responses considering the chitin particle geometry,
lattice arrangement and filling fractions up to now. Also, we analyses
the reflectance dependencies on the \(\theta\) angle that control the
helix pitch \(p\) of the helical cuticle structure through \(p=360
d/\theta\). We show how the ripples shape of \(\langle
R\rangle_{\text{bgcp}}\) \cite{Carter(2016)} can arise from unbalanced
contribution of orthogonal components when \(\theta\) is not
commensurable with some integer of the full helix rotation. We found
that shifts to blue of \(\langle R\rangle_{\text{bgcp}}\) proportionally
to \(\theta\) increases as we can expect from the decrease of \(p\) and
Eq.\eqref{eq:bragg} promoting structural color to changes from red to
blue by relatively smalls changes in \(\theta\).

\section*{Acknowledgment}
\label{sec:orgb385a40}
GPO thanks to SGCyT-UNNE for financial support through grant PI18F008.
WLM thanks DGAPA-UNAM for support through grant IN109822.

\bibliography{referencias}

\begin{thebibliography}{10}

\bibitem{Neville(1969)}
A.C.Neville and S.Caveney.
\newblock Scarabaeid beetle exocuticle as an optical analogue of cholesteric
  liquid crystals.
\newblock {\em Biol. Rev. Camb. Philos Soc.}, 44:531--562, 1969.

\bibitem{Parker(1998)}
Andrew~R. Parker, David~R. Mckenzie, and Maryanne C.~J. Large.
\newblock {Multilayer reflectors in animals using green and gold beetles as
  contrasting examples}.
\newblock {\em Journal of Experimental Biology}, 201(9):1307--1313, 05 1998.

\bibitem{Galusha(2008)}
Jeremy~W. Galusha, Lauren~R. Richey, John~S. Gardner, Jennifer~N. Cha, and
  Michael~H. Bartl.
\newblock Discovery of a diamond-based photonic crystal structure in beetle
  scales.
\newblock {\em Phys. Rev. E}, 77:050904, May 2008.

\bibitem{Sharma(2009)}
Vivek Sharma, Matija Crne, Jung~Ok Park, and Mohan Srinivasarao.
\newblock Structural origin of circularly polarized iridescence in jeweled
  beetles.
\newblock {\em science}, 325(5939):449--451, 2009.

\bibitem{Neville(1975)}
Anthony~C. Neville.
\newblock {\em Biology of the Arthropod Cuticle}.
\newblock Spring-Verlag, 1975.

\bibitem{Luna(2013)}
Ana Luna, Demetrio Mac\'{i}as, Diana Skigin, Marina Inchaussandague, Daniel
  Schinca, Miriam Gigli, and Alexandre Vial.
\newblock Characterization of the iridescence-causing multilayer structure of
  the ceroglossus suturalis beetle using bio-inspired optimization strategies.
\newblock {\em Opt. Express}, 21(16):19189--19201, Aug 2013.

\bibitem{Gullan(2014)}
P.J. Gullan and P.S. Cranston.
\newblock {\em The Insects: An Outline of Entomology}.
\newblock J. Wiley \& Sons, Ltd., Oxford, UK, fifth edition, 2014.

\bibitem{Parker(2000)}
Andrew~Richard Parker.
\newblock 515 million years of structural colour.
\newblock {\em Journal of Optics A: Pure and Applied Optics}, 2(6):R15, nov
  2000.

\bibitem{Lenau(2008)}
T.~Lenau and M.~Barfoed.
\newblock Colours and metallic sheen in beetle shells — a biomimetic search
  for material structuring principles causing light interference.
\newblock {\em Advanced Engineering Materials}, 10(4):299--314, 2008.

\bibitem{Neville(1993)}
Anthony~Charles Neville.
\newblock {\em Biology of Fibrous Composites: Development beyond the Cell
  Membrane}.
\newblock Cambridge University Press, 1993.

\bibitem{Pye(2010)}
J.~David Pye.
\newblock The distribution of circularly polarized light reflection in the
  scarabaeoidea (coleoptera).
\newblock {\em Biological Journal of the Linnean Society}, 100(3):585--596,
  2010.

\bibitem{Carter(2016)}
I.~E. Carter, K.~Weir, M.~W. McCall, and A.~R. Parker.
\newblock Variation in the circularly polarized light reflection of {Lomaptera}
  ({Scarabaeidae}) beetles.
\newblock {\em Journal of The Royal Society Interface}, 13(120), 2016.

\bibitem{Bouligand(1972)}
Y.~Bouligand.
\newblock Twisted fibrous arrangements in biological materials and cholesteric
  mesophases.
\newblock {\em Tissue and Cell}, 4(2):189--217, 1972.

\bibitem{Jewell(2007)}
Sharon~A. Jewell, Peter Vukusic, and Nicholas~W. Roberts.
\newblock Circularly polarized colour reflection from helicoidal structures in
  the beetle plusiotis boucardi.
\newblock {\em New Journal of Physics}, 9:99 -- 99, 2007.

\bibitem{Hecht(1986)}
E.Hecht and A.Zajac.
\newblock {\em Optica}.
\newblock Addison-Wesley Iberoamericana, S.A., 1986.

\bibitem{McCall(2015)}
M.W. McCall, I.J. Hodgkinson, and Q.~Wu.
\newblock {\em Birefringent Thin Films and Polarizing Elements}.
\newblock Imperial College Press, London, second edition, 2015.

\bibitem{Oldenbourg(1989)}
Rudolf Oldenbourg and Teresa Ruiz.
\newblock Birefringence and macromolecules wiener's theory revisited, with
  application to dna and tabacco mosaic virus.
\newblock {\em Biophys. J.}, 56:195--205, 1989.

\bibitem{Mochan(2016)}
W.~Luis Moch\'an, Guillermo Ortiz, Bernardo~S. Mendoza, and Jos\'e~Samuel
  P\'erez-Huerta.
\newblock Photonic.
\newblock Comprehensive Perl Archive Network (CPAN), 2016.
\newblock Perl package for calculations on metamaterials and photonic
  structures.

\bibitem{BornWolf(1999)}
M.Born and E.~Wolf.
\newblock {\em Principles of Optics}.
\newblock Cambridge University Press, seven edition, 1999.

\bibitem{Missoni(2020)}
Leandro~L. Missoni, Guillermo~P. Ortiz, María~Luz {Martínez Ricci}, Victor~J.
  Toranzos, and W.~Luis Mochán.
\newblock Rough 1d photonic crystals: A transfer matrix approach.
\newblock {\em Optical Materials}, 109:110012, 2020.

\bibitem{Puente(2020)}
Luis~Eduardo Puente-Díaz, Victor Castillo-Gallardo, Guillermo~P. Ortiz,
  José~Samuel Pérez-Huerta, Héctor Pérez-Aguilar, Vivechana Agarwal, and
  W.~Luis Mochán.
\newblock Stable calculation of optical properties of large non-periodic
  dissipative multilayered systems.
\newblock {\em Superlattices and Microstructures}, 145:106629, 2020.

\bibitem{Yeh(2005)}
Pochi Yeh.
\newblock {\em Optical Waves in Layered Media}.
\newblock Wiley Series in Pure and Applied Optics. Wiley, Hoboken, New Jersey,
  USA, 2005.

\end{thebibliography}
\end{document}